\documentclass[namedreferences]{solarphysics}
\usepackage[hyperref,optionalrh,solaromanenum]{spr-sola-addons} 
\usepackage{graphicx}                    
\usepackage{amssymb}                    
\usepackage{color}                       
\usepackage{epstopdf}
\usepackage{verbatim}
\usepackage{listings}
\usepackage{natbib}
\usepackage{multirow}
\usepackage{bigstrut}
\usepackage{booktabs}
\usepackage{float}
\usepackage{epigraph}
\usepackage{rotating}
\usepackage{longtable}
\usepackage{subfig}
\usepackage{tabularx}
\usepackage{bm}
\usepackage{array}

\newcolumntype{C}[1]{>{\centering\let\newline\\\arraybackslash\hspace{0pt}}m{#1}}
\newenvironment{dpcis}[1][Disclosure of Potential Conflicts of Interest]{\footnotesize\paragraph*{#1}}{}

\begin{document}

\begin{article}

\begin{opening}

\title{Topology of Coronal Magnetic Fields: Extending the Magnetic Skeleton Using
Null-like Points}

\author[addressref={uclan},corref,email={DLee9@uclan.ac.uk}]{\inits{D.T.}\fnm{Daniel T.}~\lnm{Lee}\orcid{0000-0002-6511-9809}}

\author[addressref={uclan},corref,email={DSBrown@uclan.ac.uk}]{\inits{D.S.}\fnm{Daniel}~\lnm{Brown}\orcid{0000-0002-1618-8816}}

\runningauthor{D. T. Lee and D. S. Brown}

\runningtitle{Extending the Magnetic Skeleton Using Null-like Points}

\address[id={uclan}]{Jeremiah Horrocks Institute, University of Central Lancashire, Preston, UK}

\begin{abstract}

    Many phenomena in the Sun's atmosphere are magnetic in nature and study of the atmospheric magnetic field plays an important part in understanding these phenomena. Tools to study solar magnetic fields include magnetic topology and features such as magnetic null points, separatrix surfaces, and separators. The theory of these has most robustly been developed under magnetic charge topology, where the sources of the magnetic field are taken to be discrete, but observed magnetic fields are continuously distributed, and reconstructions and numerical simulations typically use continuously distributed magnetic boundary conditions. This article investigates the pitfalls in using continuous source descriptions, particularly when null points on the $z=0$ plane are obscured by the continuous flux distribution through, e.g., the overlap of non-point sources. The idea of null-like points on the boundary is introduced where the parallel requirement on the field $B_{\parallel}=0$ is retained but the requirement on the perpendicular component is relaxed, i.e., $B_{\perp}\ne0$. These allow the definition of separatrix-like surfaces which are shown (through use of a squashing factor) to be a class of quasi-separatrix layer, and separator-like lines which retain the x-line structure of separators. Examples are given that demonstrate that the use of null-like points can reinstate topological features that are eliminated in the transition from discrete to continuous sources, and that their inclusion in more involved cases can enhance understanding of the magnetic structure and even change the resulting conclusions. While the examples in this article use the potential approximation, the definition of null-like points is more general and may be employed in other cases such as force-free field extrapolations and MHD simulations.
    
\end{abstract}

%

\end{opening}


\section{Introduction}\label{sec:int}

    The Sun's corona is home to a variety of violent, dynamical events, often termed `solar activity'. As we further advance our arsenal of satellites in orbit about the Earth, increasing our dependence on their successful operation, we indirectly increase the effect of space weather events on our daily routines. To better prepare for, and safeguard against, space weather events, we must increase our understanding of the mechanisms which trigger them. Activity in the solar corona is dominated by the solar magnetic field, so one must consider the Sun's magnetic field whenever one wishes to think about space weather events. 
    
    When considering the Sun's magnetic field, concepts from magnetic topology, such as magnetic null points, separatrix surfaces, and separators are often employed \citep[e.g.,][]{Antiochos1998,Beveridge2002,Brown1999b,Maclean2006a}. The theory of these has been highly developed using magnetic charge topology, which was first presented by \citet{Baum1980} and later termed \textit{`magnetic charge topology'} by \citet{Longcope1996b}. This typically models solar magnetic fields using the potential field approximation arising from point-source configurations. However, many practical investigations \citep[e.g.,][]{Aulanier2005a,Kumar2018,Regnier2008} of solar magnetic fields use continuous source configurations, either observational (e.g., from line-of-sight magnetograms), numerical (with applied photospheric boundary conditions), or both (data driven numerical simulations). Such continuous source configurations introduce a new concept for defining domains of connectivity, called a quasi-separatrix layer (QSL). QSLs are regions where the connectivity of the magnetic field changes rapidly, but continuously \citep{Mandrini1996b, Priest1995, Demoulin1996,Demoulin1997}. The effect that QSLs have on the field line mappings can be quantified \citep{Titov1999c}, is of great import when considering the energy release of a solar flare \citep{Mandrini1996b, Demoulin1997, Titov1999b, Titov1999a}, and is particularly useful when highlighting the locations of QSLs in continuous field analysis \citep{Restante2009}.
    
    While many concepts in magnetic topology easily transfer to the continuous regime, there are potential pitfalls and issues that may occur when doing so. This article looks at one of these, the loss of photospheric null points in moving from a discrete to a continuous description. When this occurs, features such as separatrix surfaces and separators that arise from the null points will also disappear from the topology, but related x-line features may still remain undetected in the configuration. To retain these features, this article introduces the idea of a null-like point (NLP, Section~\ref{sec:nlp}) which relaxes some of the restrictions of a null point but may still produce separatrix- and separator-like structures. A simple demonstration of when an NLP may occur is given, and the similarities to null points illustrated.
    
    A more involved topology with four sources is given in Section~\ref{sec:csset}. This compares the discrete and continuous cases and shows that when null points disappear between the two cases, the null-like point can be used to recover the key topological features. This is also used to demonstrate that separatrix-like surfaces associated with null-like points are actually a class of quasi-separatrix layers.
    
    Finally, Section~\ref{sec:oss} presents the example of an open-separatrix surface, as illustrated by \citet{Priest2014}. This example demonstrates some of the pitfalls of simply transferring discrete topology structures to a continuous description. It is shown that with the inclusion of null-like points and the associated separatrix-like surfaces, the so-called open separatrix is not as open as it initially appears! 
    
    \section{Magnetic Field Models}
    
    \subsection{Magnetic Charge Topology}
    
    The magnetic charge topology (MCT) is used in this article for comparison to continuous cases. MCT makes two approximations in order to calculate the topology of a magnetic field. First, the coronal field is assumed to be potential, which is a particular solution of a force-free field. This may be a valid assumption, provided the velocities in the plasma are much smaller than the Alfv\'{e}n speed. The potential field solution then produces a topology which is robust to small  non-potential perturbations to the field, so the field may be considered to evolve through a series of static equilibria. This first approximation is feasible in quiet-Sun regions, but almost certainly breaks down during the eruptive phase of a solar flare. 

    Secondly, magnetic sources are described by discrete points upon the photospheric plane ($z = 0$), where each source has a radius of zero. This approach is used as an approximation of continuously distributed sources; at the solar surface, discrete points of magnetic flux are not observed. However, provided a continuous source has a sufficiently small radius, an observer at a large distance may perceive the source to be a single point. At the point sources themselves, $\bm{\nabla}~.~\mathbf{B} = 0$, but $\mathbf{B}$ becomes infinite in violation of the divergence theorem. This problem is resolved by placing the sources on a bounding surface of the volume modelled. These surfaces, or at least the points local to the point source, can then be removed from the volume, hence $\bm{\nabla}~.~\mathbf{B} = 0$ is preserved everywhere in the simulated volume. Together, these assumptions allow the magnetic field to be calculated analytically at any point in the volume. It can be described by the equation, 
    \begin{equation}\label{eqn:magsum}
        \mathbf{B}(\mathbf{r}) = \sum_{i}^{} \varepsilon_i \frac{\mathbf{r} - \mathbf{r}_i}{\mid \mathbf{r} - \mathbf{r}_i \mid ^3}, 
    \end{equation}
    where, $\mathbf{r} = x \mathbf{\hat{x}} + y \mathbf{\hat{y}} + z \mathbf{\hat{z}}$ and $\mathbf{r}_i = x_i \mathbf{\hat{x}} + y_i \mathbf{\hat{y}} + z_i \mathbf{\hat{z}}$ is the position of source $i$ with flux $\varepsilon_i$. Thus, a magnetic topology can be characterized using this equation and a particular source configuration, as done by \citet{Brown1999a} when producing the complete classification of magnetic topologies generated by three discrete sources. Where a source configuration has an imbalance of magnetic flux on the photospheric plane, a balancing source is assumed to exist at infinity \citep{Inverarity1999}.
    
    \subsection{Continuous Source Potential Fields}\label{sec:cspf}
    
    Magnetic sources at the photosphere are not observed as discrete points, but as continuously distributed flux regions. It may not always be appropriate to assume a source to be a discrete point, such as the large concentrations observed within an active region. The continuous-source model addresses this by assuming each source lies within the photospheric plane, $z = 0$, and its flux is distributed across a shape.
    
    Unlike the point source model, the solution for a magnetic field in the continuous source regime is not (usually) analytic. Instead, the magnetic field must be calculated by defining a magnetic scalar potential with,
    \begin{equation}
        \mathbf{B}=\nabla\psi,
    \end{equation}
    and numerically solving Poisson's equation,
    \begin{equation}
        \nabla^{2}\psi = 0,
    \end{equation}
    across the volume subject to an imposed normal component of the magnetic field at the boundary of the numerical domain. An alternating direction-implicit (ADI) method \citep{Morton2005} is then employed to solve Poisson's equation in each of the axes across a staggered grid. The three dimensional volume is then considered in a series of one-dimensional layers consisting of grid squares. To set up the boundary conditions, the $B_{z}$ component in the $z=0$ plane is imposed according to the desired source setup. There are then three choices for the side and top boundaries: (a) if the sources are balanced then the remaining normal components can be set to 0, (b) if the sources are unbalanced the imbalance can be distributed equally across the top boundary, and (c) in either case the normal components can be approximated using Equation~\ref{eqn:magsum} where each grid cell is treated as a point source and then the entire boundary flux is scaled by an amount to ensure flux balance. Option (c) is best used when comparing with discrete setups, but options (a) and (b) can be more practical when investigating the squashing factor (Section~\ref{sec:SFINLP}).
    
    
    \subsection{Quasi-Separatrix Layers}
    
    Whilst MCT provides a computationally cheap means of modelling magnetic topologies, it does not include all of the topological features that are present. For example, a point source model cannot have two sources partially overlap, whereas a continuous source model can.
    
	Another feature `missing' from the MCT approach was identified in studies of solar flares, which found that the separatrix surfaces of magnetic null points are not always associated with the flaring site \citep{Demoulin1994}. However, there was a strong correlation between these sites and steep gradients in the field line connectivities \citep{Mandrini1995, Demoulin1997}, and \citet{Priest1995} later provided a definition for these regions, naming them quasi-separatrix layers. This definition is subsequently adopted by \citet{Mandrini1996b,Mandrini1996a} to identify quasi-separatrix layers in X-ray bright points.
	
	Quasi-separatrix layers (QSLs) are an important feature of the magnetic skeleton, where the connectivity of field lines changes rapidly over small scales but is not discontinuous as at a separatrix surface \citep{Priest1995, Demoulin1996,Mandrini1996b,Demoulin1997}. QSLs intersect one another at quasi-separators, which provide locations for reconnection to occur in a similar manner to separators \citep{Priest2014}. However, QSLs are regions where reconnection can occur in the absence of separatrices, separators, and null points \citep{Priest1995}, hence it is important to know if they are present in a given topology, particularly in the context of energy release in a solar flare \citep{Demoulin1997, Titov1999b, Titov1999a}.
	
    \citet{Priest1995} defined a function, $N(\mathbf{r}_{\pm})$, of the footpoints of closed magnetic field lines, to identify QSLs as regions where $N(\mathbf{r}_{\pm}) \gg 1$. However, this result varied such that one field line may be attributed two values for $N$, dependent on the direction of the mapping from one footpoint to the other. \citet{Titov1999c} refined this definition of a QSL by calculating a squashing factor, $Q$, which described the mappings in terms of their derivatives. Hence, $Q$ characterises the magnetic connectivity of the volume, rather than the field line mappings. Unlike the function $N(\mathbf{r}_{\pm})$, the definition for $Q$ is invariant, such that the same value is retrieved regardless of the direction of the mapping \citep{Titov1999c}. 
    
    The definition of the squashing factor, $Q$, has been applied to other scenarios in a planar geometry \citep{Titov2002b, Titov2002a} and has been generalised for a global geometry \citep{Titov2007}. In this article, $Q$ will be used under the principles of MCT to aid the interpretation of simulated magnetic topologies. 
    
    Given the importance of QSLs in the reconnection process, it is `useful' to also include them in topological models of the Sun's magnetic field. Currently, an MCT approach makes no such allowance, but the inclusion of null-like points provides a straight-forward way of locating a class of quasi-separatrix layer in continuous source configurations. 

\section{Defining Null-like Points (NLP)}\label{sec:nlp} 

    Under the principles of MCT, if two point sources of the same polarity are close together, there is usually a null point located between them. Linear magnetic null points were first defined by \citet{Lau1990} as regions where the magnetic field vanishes ($\mathbf{B} = 0$). In three dimensions, this means, 
    \begin{equation}\label{eqn:bnp}
    B_{x} = B_{y} = B_{z} = 0,
    \end{equation}
    where $B_{x}, B_{y}$, and $B_{z}$ are the components of the magnetic field in the $x$, $y$, and $z$ directions respectively. The spine and fan field lines of a null point are found from the Jacobian matrix of the linearised field local to the null point \citep{Priest1996}, and a null point can be described as prone or upright, depending on the orientation of its spine field lines \citep{Maclean2005}. However, a point source model may not always be appropriate, such as when modelling a source with a complex shape, or a source region with multiple overlapping sources. 
    
    \begin{figure}[t]
        \centering
        \subfloat[]{\includegraphics[width=0.5\textwidth]{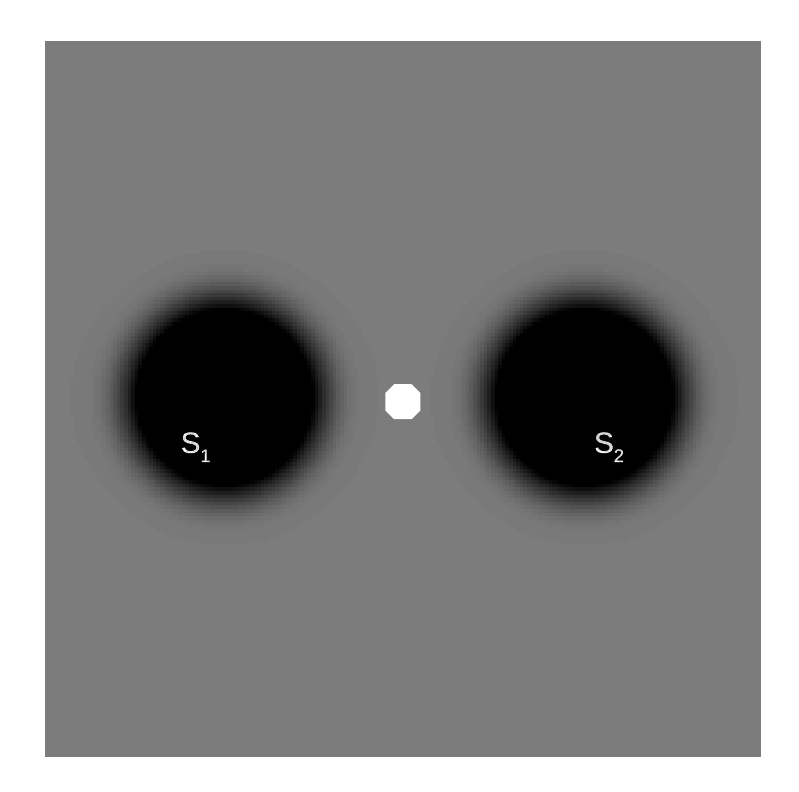}\label{fig:nlpex5}}
        \hfill
        \subfloat[]{\includegraphics[width=0.5\textwidth]{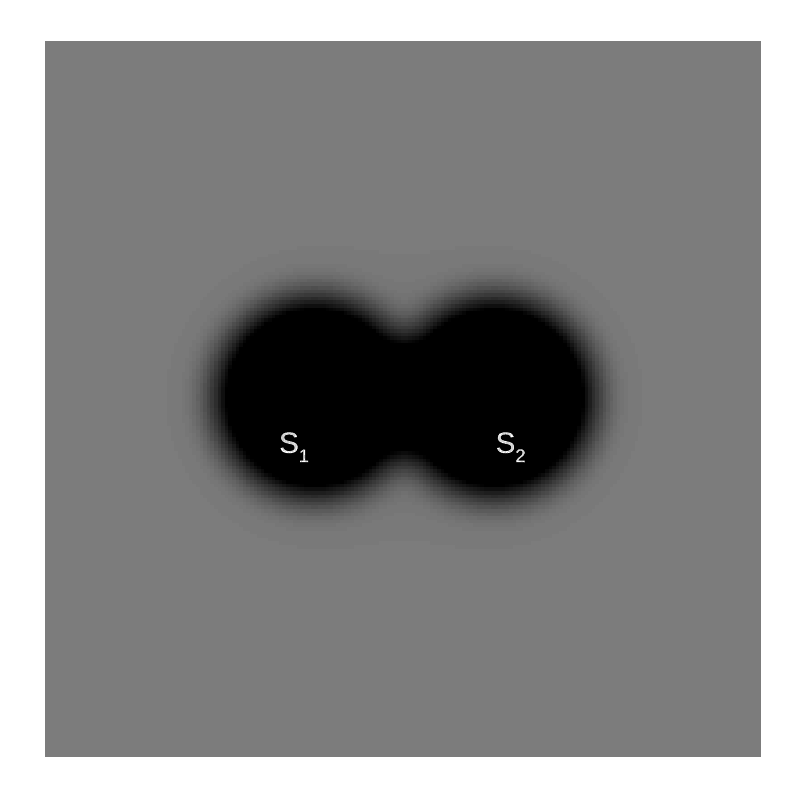}\label{fig:nlpex6}}\\
        \subfloat[]{\includegraphics[width=0.5\textwidth]{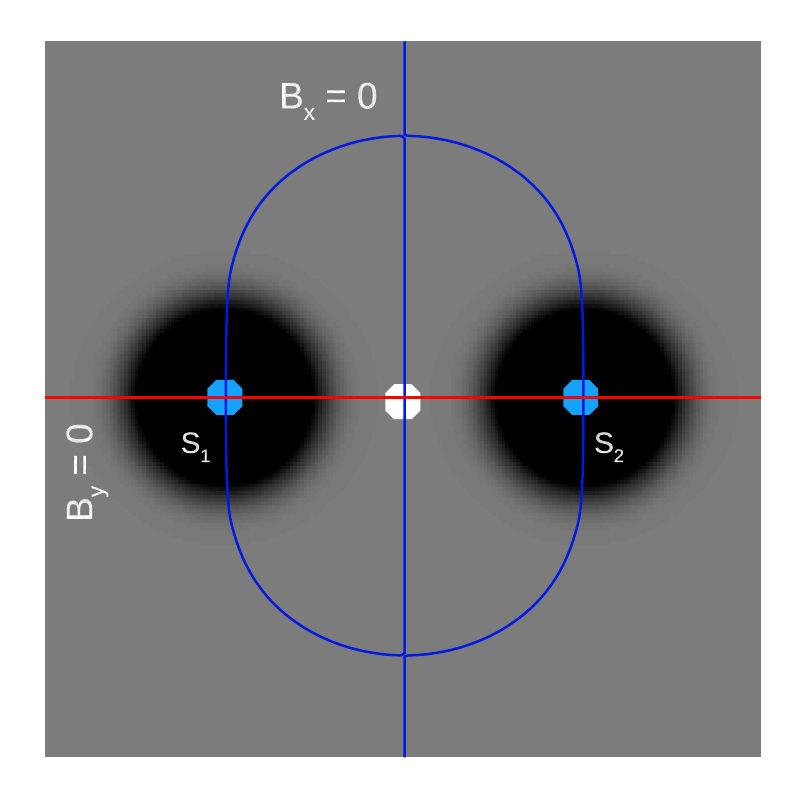}\label{fig:nlpex1}}
        \hfill
        \subfloat[]{\includegraphics[width=0.5\textwidth]{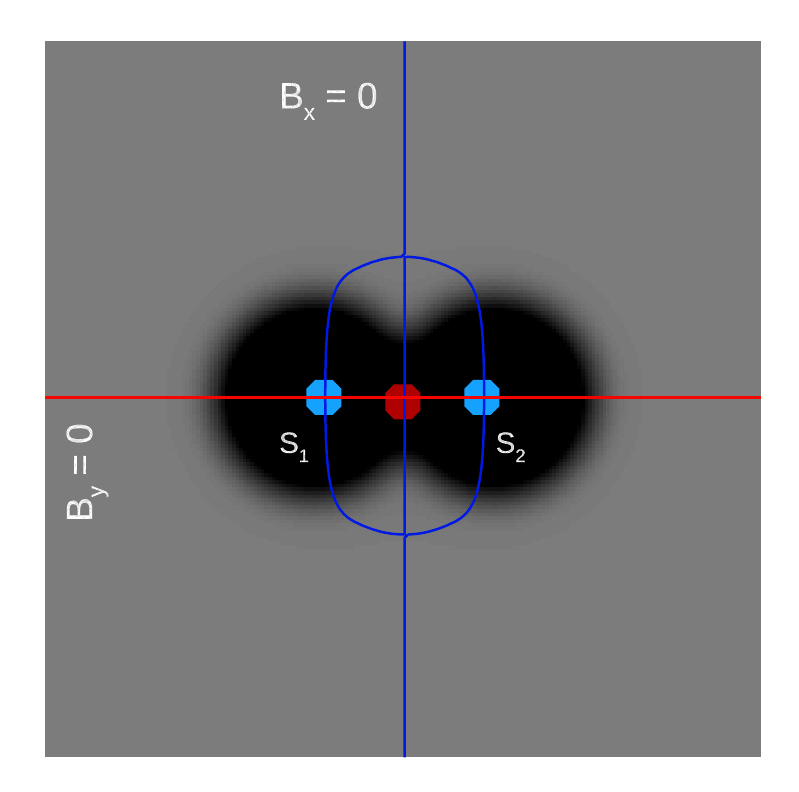}\label{fig:nlpex2}}\\
        \caption{a) Planar plot of two separate negative sources of flux, where the white dot indicates the position of the null point. b) Planar plot of two circular sources of flux which now overlap, where the previous null has now vanished. c) The same configuration as in panel a, with nullclines plotted to show the x-points in the $x$-$y$ plane. The blue line gives the $B_x$ nullcline and the $B_y$ nullcline is shown in red. Source-like points have also been located and plotted as blue dots. d) The same configuration as panel b, though the $B_x$ and $B_y$ nullclines still intersect at the point where the null was previously located, with the located NLP plotted as a red dot.}\label{fig:nlpexa}
    \end{figure}
    The continuous source model adds the ability to better represent these complexities. Similar to the MCT approach, placing two small sources of the same polarity close together, produces a null point between them (Figure \ref{fig:nlpex5}). However, when the two sources of magnetic flux overlap, the null point between them vanishes (Figure \ref{fig:nlpex6}) as Equation \ref{eqn:bnp} no longer holds ($B_z \neq 0$). In the absence of the null point, an x-line structure may persist within the overlap of the two sources. 

    Within the boundary of a continuously distributed source, the magnetic field will never vanish completely. Thus, to locate these x-line structures, the definition of a magnetic null point is relaxed such that, 
    \begin{eqnarray}
        B_{x} = B_{y} & = & 0,\label{eqn:bxynlp} \\
        B_{z} & \neq & 0. \label{eqn:bznlp}
    \end{eqnarray}
    There are then three cases to consider, regarding the eigenvalues $\mu_1$ and $\mu_2$ of the reduced Jacobian matrix in the horizontal plane of the magnetic field components in the $B_x$ and $B_y$ system only, along with the sign of $B_z$. 
    
    \begin{figure}[t]
        \centering
        \subfloat[Prone NLP]{\includegraphics[width=0.3\textwidth]{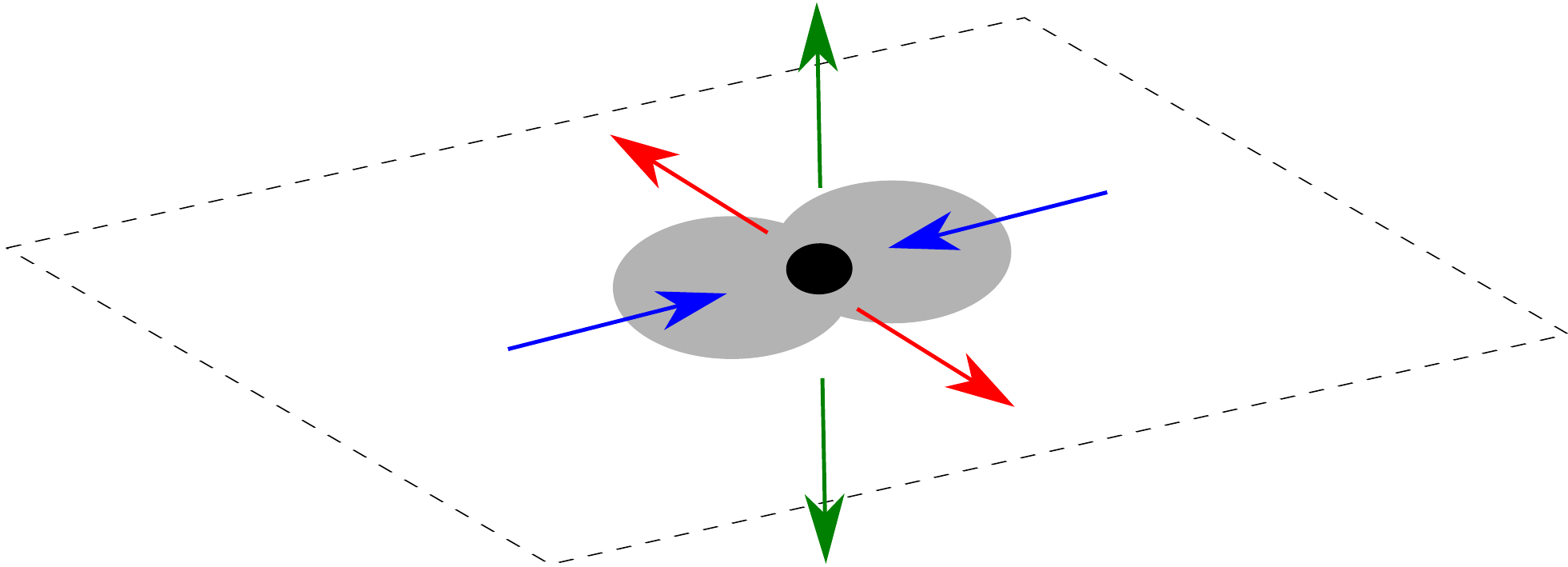}\label{fig:pnlpsketch}}
        \subfloat[Upright NLP]{\includegraphics[width=0.3\textwidth]{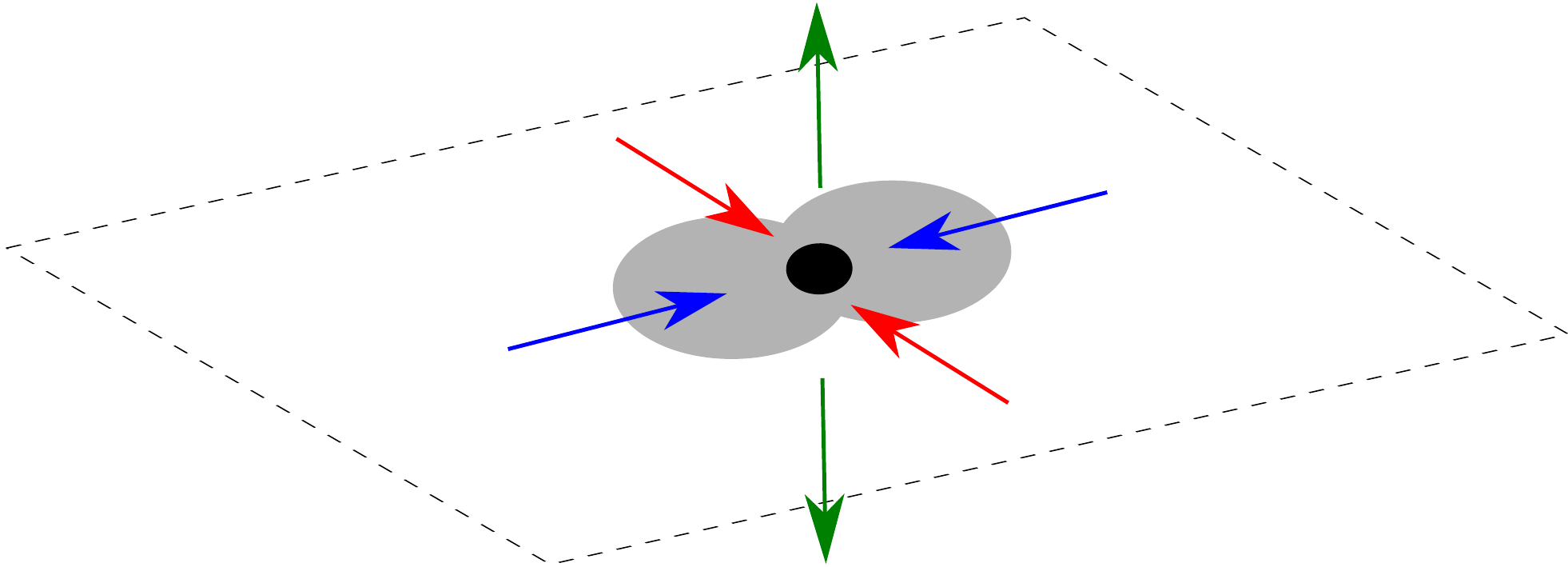}\label{fig:unlpsketch}}
        \subfloat[SLP]{\includegraphics[width=0.3\textwidth]{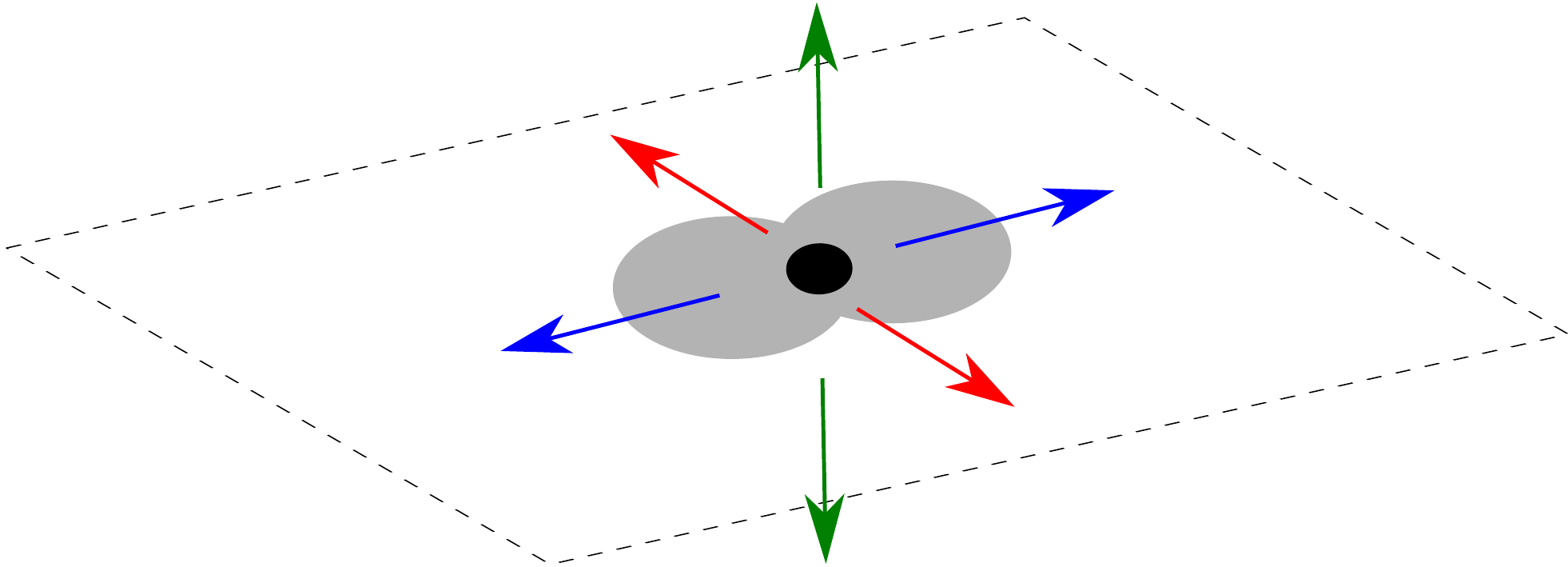}\label{fig:slpsketch}}\\
        \caption{Illustration of the three possible configurations located on a continuous source flux distribution (grey circles). Green arrows represent the direction of the $B_z$ component of the magnetic field at an identified point, which is positive in this case. The blue and red arrows represent the direction of the eigenvectors corresponding to $\mu_1$ and $\mu_2$. The three configurations give: a) a representation of a prone NLP, whose separatrix-like surface would project in the $y-z$ plane, b) a representation of an upright NLP, whose spine-like field lines would propagate in the $z$ direction, and c) a representation of an SLP acting as a source. Mirror versions of each of these where the direction of all arrows are reversed are also possible.} \label{fig:nlpsketch}
    \end{figure}

    Suppose first that $\mu_1$ and $\mu_2$ are of opposite sign. One of these eigenvalues, say $\mu_2$, will have a sign aligned with the direction of the $B_z$ component. This configuration is sufficient to preserve the x-point configuration in the $x$-$y$ plane. Here, a location where these conditions are met is called a prone null-like point (NLP), serving as the signature of an x-line structure in the photospheric plane (Figure \ref{fig:pnlpsketch}).

    The fan field lines of an NLP form a separatrix-like surface. To calculate field lines in this separatrix-like surface, the two-dimensional separatrix in the $z=0$ plane aligned with the eigenvector $\bm{v}_2$ associated with $\mu_2$ is calculated. The start points of field lines are chosen along this separatrix where it contained within the continuous source. Such a separatrix-like surface is a QSL, though not all QSLs will be separatrix-like surfaces. The intersection of a separatrix-like surface with another separatrix surface produces a separator-like field line, while the intersection of two QSLs produces a quasi-separator \citep[as defined in][]{Priest2014}. Spine field lines of prone NLPs are not worth calculating since they terminate immediately as the NLP is located in the spine-terminating source region. 

    Next, consider the case where the eigenvalues $\mu_1$ and $\mu_2$ are of the same sign, but the $B_{z}$ component has opposite sign. A location where these conditions hold is defined as an upright NLP, where the associated separatrix-like surface is bound to the photospheric plane, and the spine-like field lines propagate out into the coronal volume (Figure \ref{fig:unlpsketch}). 

    The final scenario is where the eigenvalues $\mu_1$, $\mu_2$, and the sign of $B_z$ are all the same. This situation may arise at local absolute maxima in the flux distribution across a continuous source. Here, these features are called source-like points (SLPs), and act as a two-dimensional source or sink in $B_{x}$ and $B_{y}$, whilst being non-stationary in $B_{z}$ (Figure \ref{fig:slpsketch}). They have similarities with point sources. 
    
    Therefore, a null-like point (NLP) is defined as a location on the boundary where:
        \begin{enumerate}
        \item Equations \ref{eqn:bxynlp} and \ref{eqn:bznlp} hold.
        \item The sign of the eigenvalues $\mu_1$, $\mu_2$, and the sign of $B_z$ are not all the same.
    \end{enumerate}
    These further sub-divide as:
    \begin{itemize}
        \item A prone NLP if $\mu_1$ and $\mu_2$ are of opposite sign;
        \item An upright NLP if $\mu_1$ and $\mu_2$ are both of opposite sign to $B_z$.
    \end{itemize}
    
    Some of these topological features are illustrated by the source configurations shown in Figures \ref{fig:nlpex1} and \ref{fig:nlpex2}. In Figure \ref{fig:nlpex1}, the two sources are distinct from one another, and a null point is found between them. Another useful topological feature is the nullcline, which is a line where one component of the magnetic field is equal to zero (i.e. where $B_x = 0$ or $B_y = 0$). The crossing of nullclines indicate (but do not discriminate between) the locations of null points, NLPs, sources, and SLPs. The nullclines for this configuration have been plotted over the region, indicating the presence of two SLPs close to the centre of the sources. In Figure \ref{fig:nlpex2}, the location of the null point has now been covered by the overlapping sources. However, the nullclines still intersect at the point where the null point was located, which now indicates the presence of an NLP. The two SLPs that where previously found are still present, though their position has moved with the position of the sources. 
    
    \begin{figure}[t]
    \centering
    \subfloat[]{\includegraphics[width=0.5\textwidth]{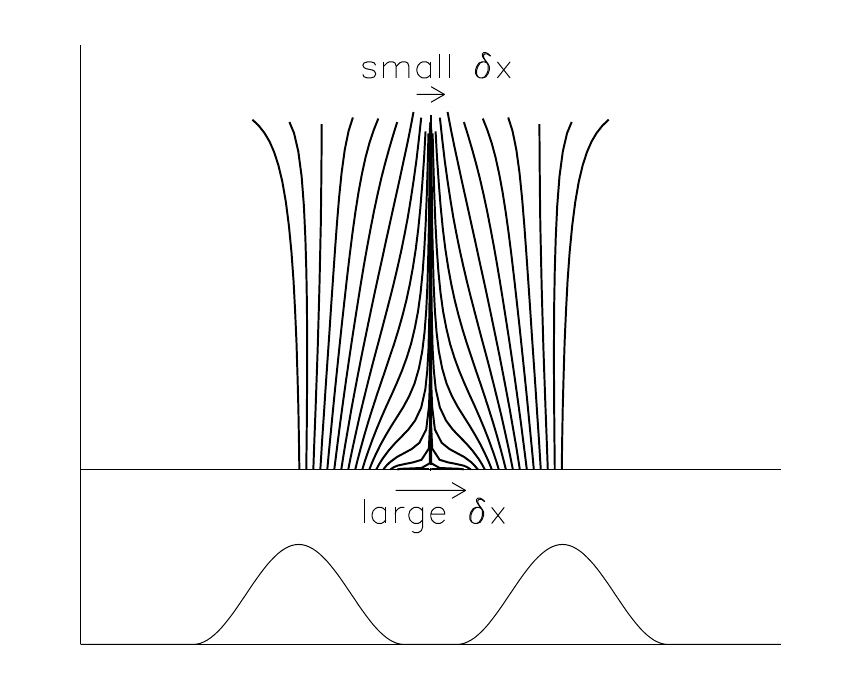}\label{fig:nlppr1}}
    \hfill
    \subfloat[]{\includegraphics[width=0.5\textwidth]{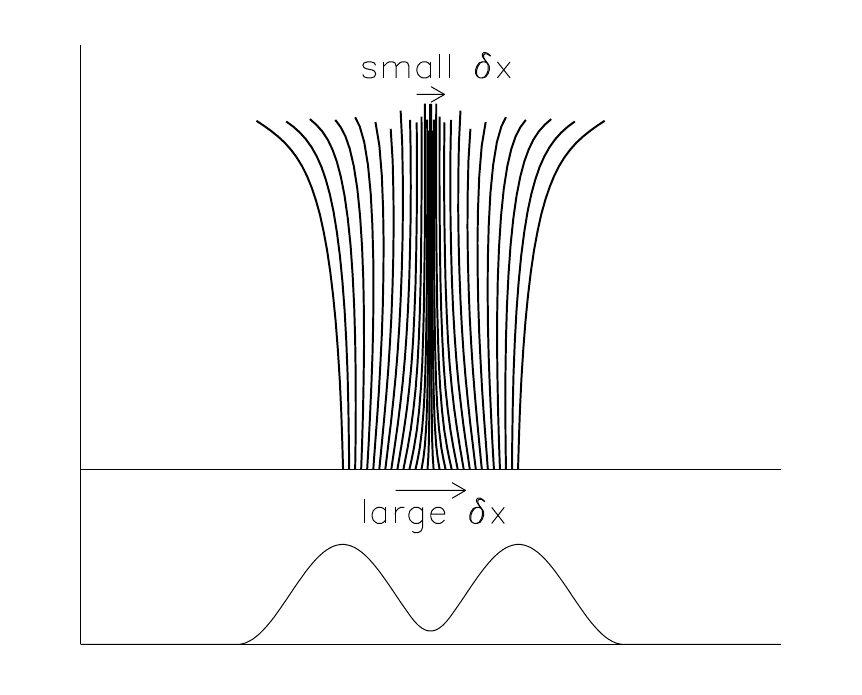}\label{fig:nlppr2}}
    \caption{a) Schematic showing the field line spread produced from field lines with footpoints on the continuous sources, when positioned apart as in Figure \ref{fig:nlpex1}. The flux distribution along the x-axis is shown in the same frame, below the field line locations. b) Displays the same information as panel a, but for the configuration where the sources overlap as shown in Figure \ref{fig:nlpex2}.}\label{fig:nlppra}
    \end{figure}
    
    Figures \ref{fig:nlppr1} and \ref{fig:nlppr2} show the vertical dispersion of magnetic field lines around a null point, and around an NLP respectively, where magnetic sources have been placed as shown in Figure \ref{fig:nlpexa}. Here, the structure of the field lines has connotations with the idea of magnetic flipping \citep{Priest1992, Priest1996} in QSLs. Consider the structure of the field lines shown in Figure \ref{fig:nlppr1}. Field lines which start closer to the edges of the circular sources are initially inclined towards the separatrix surface of the null point which lies between the two sources. As the field lines get close to this surface they are deflected upwards and the separation between them is greatly reduced. This shows the discontinuous mapping of the field lines. Now consider the field lines shown in Figure \ref{fig:nlppr2}. Similar behaviour is exhibited by these field lines as they get close to the location of the NLP, and hence close to the separatrix-like surface associated with the NLP, where the field lines begin to bunch together. While the field line mapping is no longer discontinuous, there is a region of rapid change in field line connectivity, which is characteristic of QSLs \citep{Mandrini1996b, Demoulin1997}.

\section{Null-like Points with a Squashing Factor}\label{sec:csset} 

    \begin{table}[t]
	    \begin{center}
		    \begin{tabular}{C{2.5cm} C{2.5cm} C{2.5cm} C{2.5cm}}
			    \multicolumn{4}{c}{\textbf{Non-Overlapping Source Parameters}} \\
			    \textbf{Source} & \textbf{x} & \textbf{y} & $\bm{\epsilon}$ \\
			    \hline
			    S1 & ~0.40 & -0.20 & ~0.50\\
			    S2 & -0.40 & -0.21 & ~0.50\\
			    S3 & -0.01 & ~0.50 & -0.50\\
			    S4 & ~0.01 &  ~1.25 & -0.50\\
    			    \\[6pt]
	    		\multicolumn{4}{c}{\textbf{Positive Overlapping Source Parameters}} \\
	    		\textbf{Source} & \textbf{x} & \textbf{y} & $\bm{\epsilon}$ \\
	    		\hline
	    		S1$_{O}$ & ~0.18 & -0.20 & ~0.50\\
	    		S2$_{O}$ & -0.18 & -0.21 & ~0.50\\
	    		\\[6pt]
	    		\multicolumn{4}{c}{\textbf{Negative Overlapping Source Parameters}} \\
    			\textbf{Source} & \textbf{x} & \textbf{y} & $\bm{\epsilon}$ \\
	    		\hline
		    	S3$_{O}$ & -0.01 & ~0.72 & -0.50\\
			    S4$_{O}$ & 0.01 & ~1.08 & -0.50\\
    		\end{tabular}
	    	\caption{Table containing the parameters for the configuration of the discrete and continuous sources used for the four-source case study. S$_{O}$ represents sources which have been re-positioned so that they overlap one another. Here, $\epsilon$ is used to represent the flux prescribed to each source.}\label{tab:4sconfigs}
	    \end{center}
    \end{table}
    
    Consider now a more involved four source case, using the modelling assumptions for point source and continuous source regimes. This example employs four sources in the $z=0$ plane and is used to demonstrate situations where magnetic charge topology in the continuous case does not fully identify the topological structure, but the addition of NLPs recovers the missing structures when moving from the discrete case. In each configuration, two of the sources are positive and two are negative. Each magnetic source has the same absolute flux (0.5 arbitrary units), thus the cases considered here exist in a state of flux balance. Whilst such an occurrence is not typical of the solar surface, these cases provide the initial steps towards formulating a foundation for the properties of NLPs.  

    Discrete source topologies and continuous source topologies are produced under equivalent source configurations and compared to one another. The parameters for all of the sources used in this case study are shown in Table \ref{tab:4sconfigs}. Where continuous sources are used, they are centred on these points, and have their flux distributed over a radius of $r = 0.28$ about this point. Artificial symmetries in the source configurations have been removed by perturbing source positions.


\subsection{Discrete and Continuous Sources without Null-like Points}\label{sec:DS4}

\begin{figure}[t]
    \centering
    \subfloat[]{\includegraphics[width=0.5\textwidth]{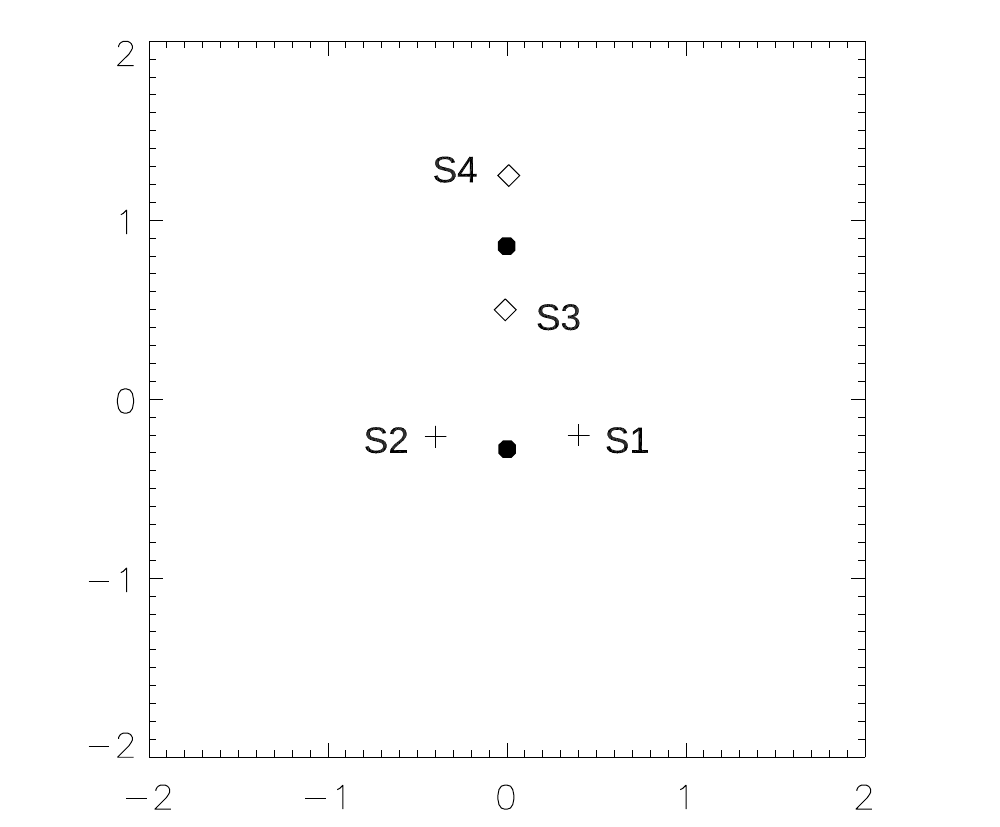}\label{fig:4dposa}}
    \hfill
    \subfloat[]{\includegraphics[width=0.5\textwidth]{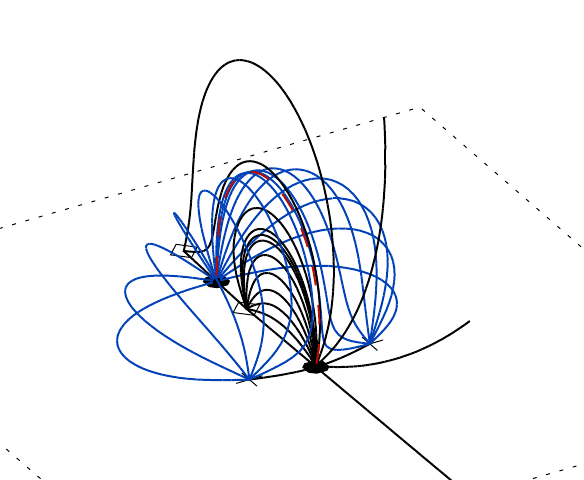}\label{fig:4dtopa}}
    \caption{Configuration and topology of the discrete, four-source case. a) The positions of the magnetic sources when they are apart from each other. Crosses mark the positions of the positive sources, diamonds are negative sources, and dots are null points. b) The magnetic skeleton produced by this source configuration, showing a separatrix dome, intersected by a separatrix wall, producing a separator.}\label{fig:fourdisci}
    \end{figure}

\begin{figure}[t]
    \centering
    \subfloat[]{\includegraphics[width=0.5\textwidth]{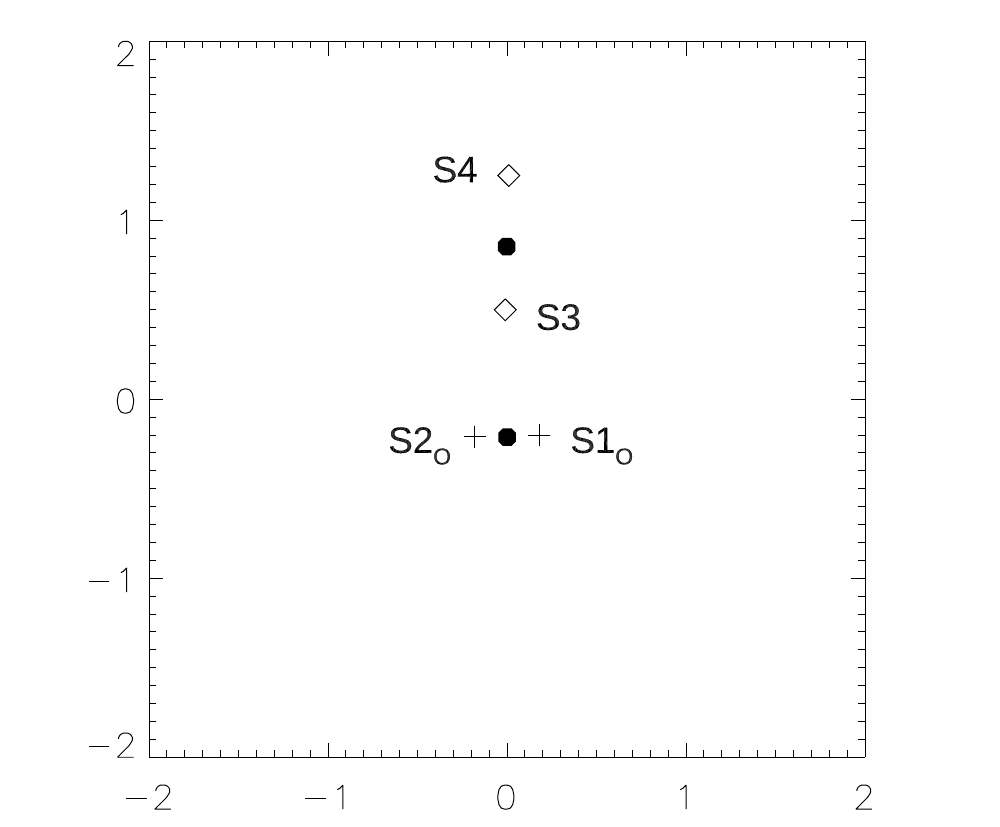}\label{fig:4dposc}}
    \hfill
    \subfloat[]{\includegraphics[width=0.5\textwidth]{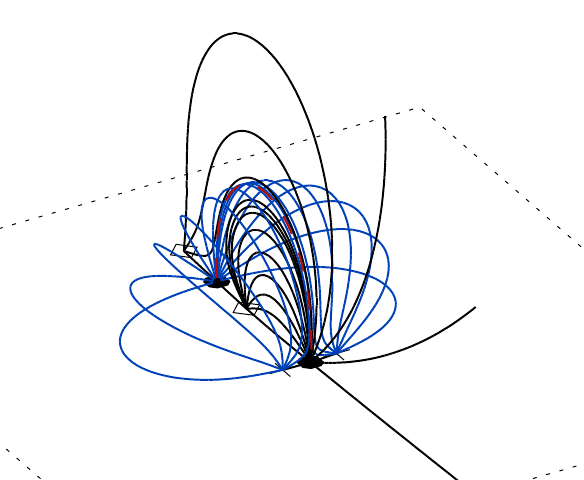}\label{fig:4dtopc}}
    \caption{a) The configuration of the sources when the positive sources are close to each other. Crosses mark the positions of the positive sources, diamonds are negative sources, and dots are null points. b) The topology produced by this configuration.}\label{fig:fourdisce}
    \end{figure}

    Initially, the four-source case is modelled by a configuration of four discrete sources of magnetic flux positioned in two source pairs (Figure \ref{fig:4dposa}), using the parameters for sources S1 - S4 shown in Table \ref{tab:4sconfigs}. A magnetic null point exists in the region between each pair of like-charged sources. The resulting topology is shown in Figure \ref{fig:4dtopa}. Here, each of the two null points give rise to a separatrix surface and a single magnetic separator exists, connecting the two null points where the separatrix surfaces intersect. This gives a topology similar to the intersected state as presented by \citet{Brown1999a}; the main difference between this topology and the three-source intersected state is the additional source of flux. The two positive sources are then positioned closer to one another (Figures \ref{fig:fourdisce} \subref*{fig:4dposc} and \subref*{fig:4dtopc}), now using sources S1$_O$ and S2$_O$ instead of S1 and S2, and the resulting topology is qualitatively the same as the topology of the first configuration. It is clear that discrete sources of flux will not `overlap' unless they are placed at the same coordinates, and so the magnetic null point between them will remain which preserves the topology of the magnetic skeleton after the positive sources are re-positioned. 


\begin{figure}[t]
    \centering
    \subfloat[First Configuration]{\includegraphics[width=0.5\textwidth]{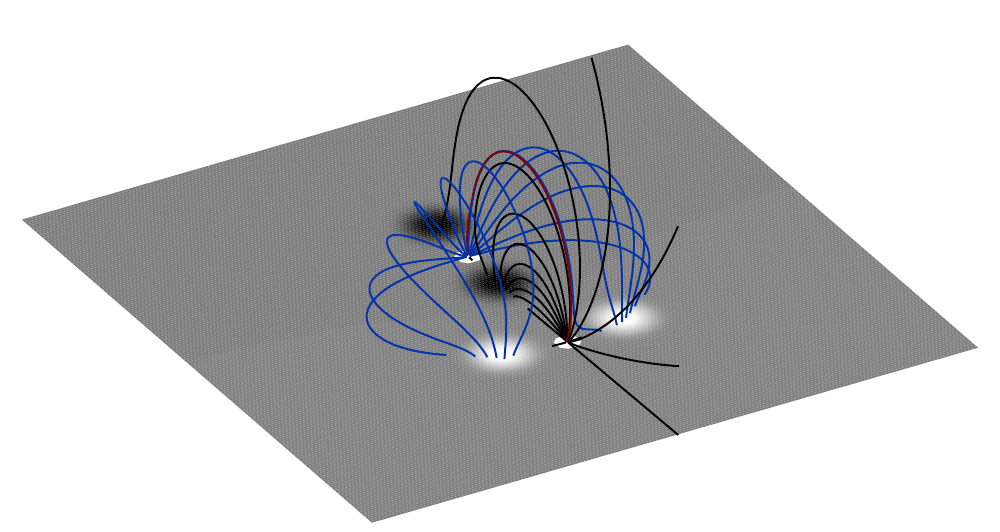}\label{fig:4ctopa}}
    \hfill
    \subfloat[Second Configuration]{\includegraphics[width=0.5\textwidth]{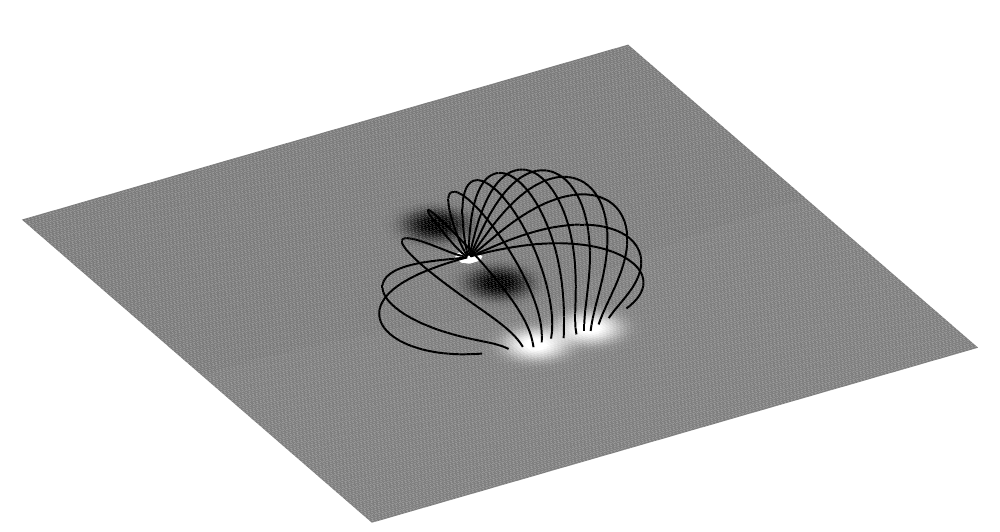}\label{fig:4ctopc}}
    \caption{A four-source topology with circular sources of magnetic flux. (a) None of the sources overlap and each pair has a null point between them. (b) The positive sources now overlap. When compared to Figure \ref{fig:4dtopc}, it is clear that the null point is lost in the overlap.} \label{fig:fourcont}
    \end{figure}

    The simulation is then repeated under the continuous source regime. In the first configuration, each of the sources is separated from each other, and the topology of the magnetic skeleton (Figure \ref{fig:4ctopa}) is the same as its counterpart in the discrete source case (Figure \ref{fig:4dtopa}). However, as the two positive sources are re-positioned closer to one another, they begin to overlap, causing the null point between them to vanish (Figure \ref{fig:4ctopc}). This changes the topology such that there is now only a separatrix dome present, where the separatrix wall and associated separator have been eliminated by the source overlap. 

\subsection{Continuous Sources with Null-like Points}

\begin{figure}[t]
    \centerline{\includegraphics[width=0.5\textwidth,clip=]{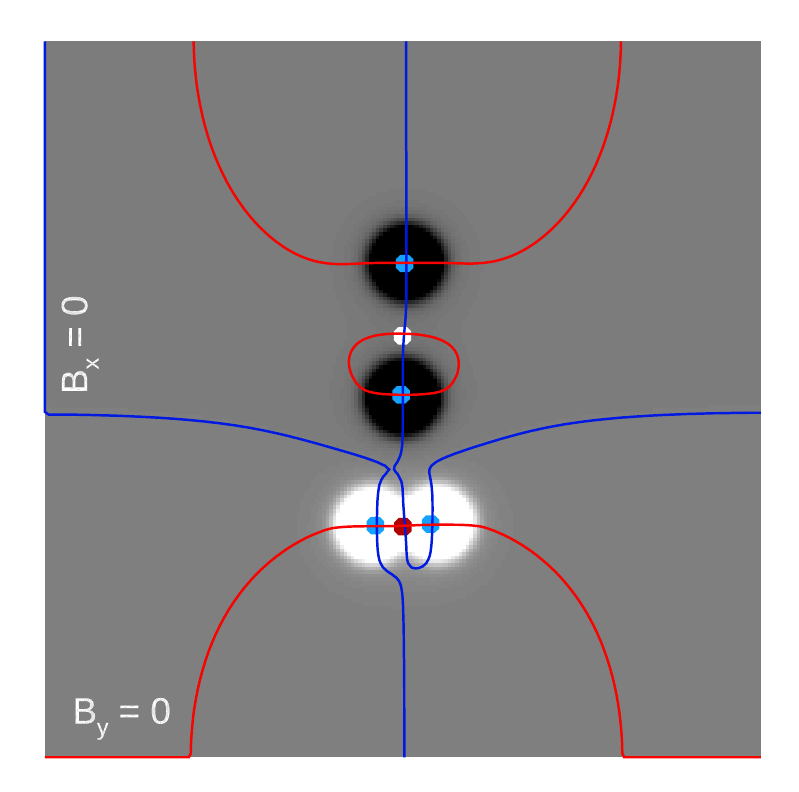}}
    \caption{Planar plot of the four-source configuration. White dots mark magnetic null points, red dots indicate NLPs and blue dots show SLPs. Nullclines are plotted as blue and red lines representing $B_x = 0$ and $B_y = 0$ respectively.}\label{fig:4snlpcomp}
    \end{figure}

\begin{figure}[t]
    \centerline{\includegraphics[width=0.5\textwidth,clip=]{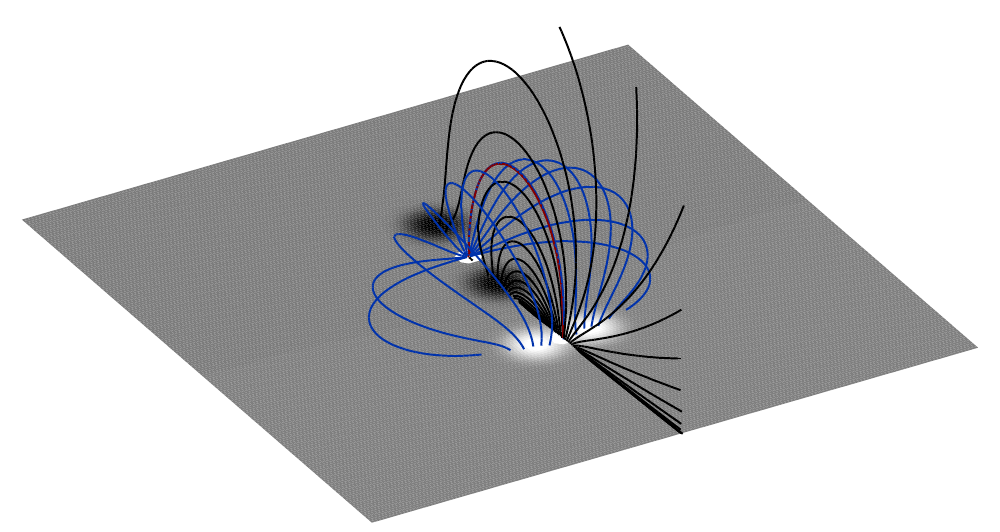}}
    \caption{A four-source topology with the positive circular sources overlapping. Here, NLPs have been included, and one has been located in the overlap of the positive sources. The combination of the separatrix and separatrix-like surface form the complete magnetic skeleton and exist in a structure analogous to that found in Section \ref{sec:DS4}.}\label{fig:4SNlP}
    \end{figure}

    The model is now adapted to include locations of both true null points and NLPs. Figure \ref{fig:4snlpcomp} shows the configuration of the magnetic sources in the $z = 0$ plane, where positive flux is shown by white regions and negative flux is represented by black regions. Nullclines where $B_x = 0$ (blue) and $B_y = 0$ (red) are plotted over the region, and where these nullclines cross indicate points where $B_x = B_y = 0$. One such location is between the two negative sources where $B_z = 0$, so a true null point is found. There are three such locations in the overlapping positive sources, two (blue) are at the source centres and correspond to SLPs. The remaining (red) location corresponds to an NLP. From this NLP, a separatrix-like surface is extended which forms a wall structure, which replicates the behaviour of the previously existing null point (figure \ref{fig:4SNlP}). This surface intersects the dome of the null between the negative sources, connecting the null to the NLP in a separator-like field line, and is co-located with an x-line structure. Thus, the topology of Figure \ref{fig:4ctopa} is preserved, using a separatrix-like surface and a separator-like field line. 

    This simplistic configuration illustrates the importance of NLPs in a continuous source regime, as they complement the magnetic skeleton produced by the magnetic sources and reveal topological-like features that might otherwise be missed.

\subsection{Associating Null-like Points with Quasi-separatrix Layers}\label{sec:SFINLP}

    Although an NLP can be located by its definition, calculation of the squashing factor, $Q$, provides further evidence for the role of the NLP in defining the separatrix-like surface. The definition of the squashing factor was presented by \citet{Titov1999c} for finding the location of QSLs. Here, it is used as an accompaniment to the definition of the NLP to demonstrate that these are locations where an x-line protrudes through the base of the simulation. 

    \begin{figure}[t]
    \centering
    \subfloat[]{\includegraphics[width=0.5\textwidth]{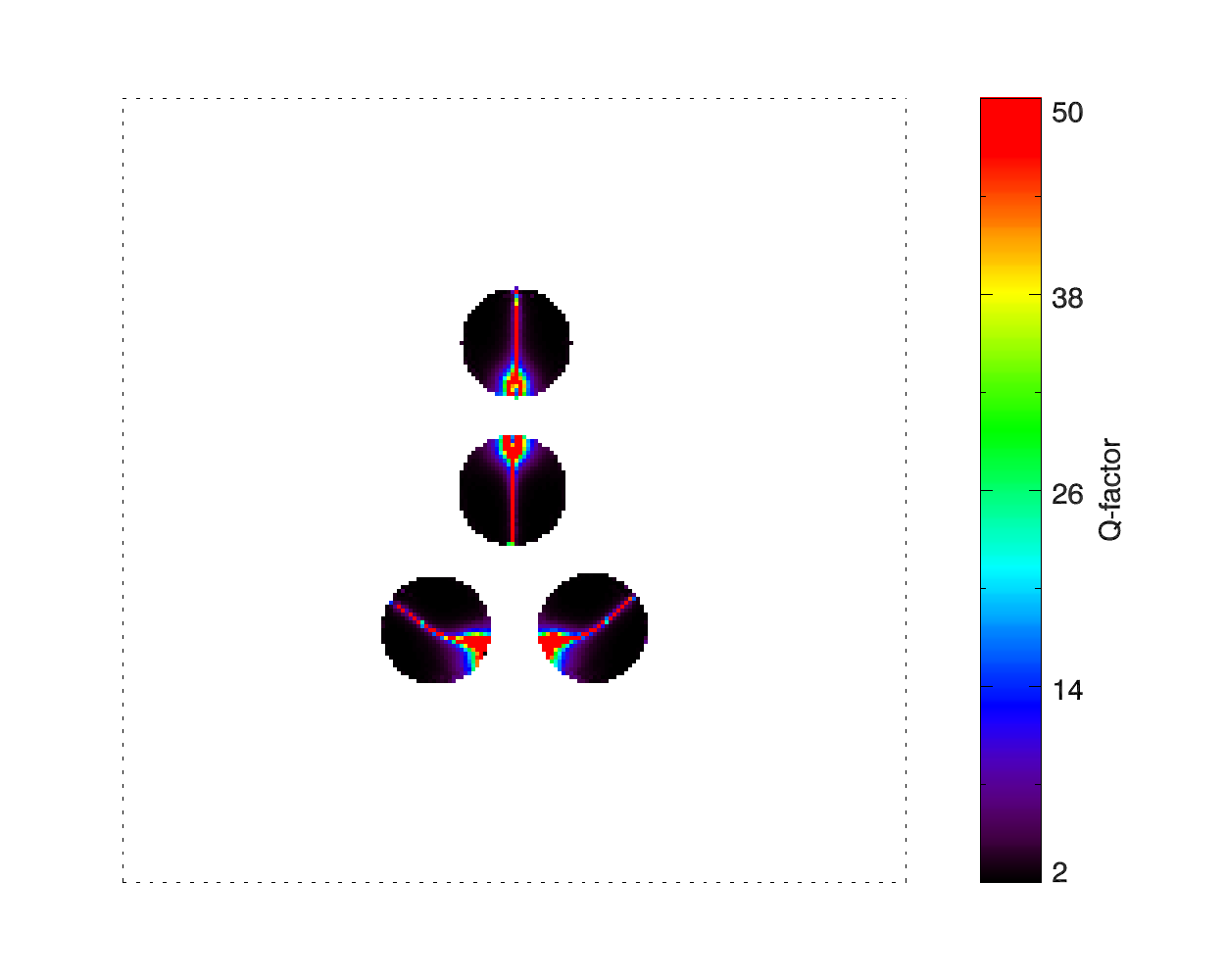}\label{fig:4snoover}}
    \hfill
    \subfloat[]{\includegraphics[width=0.5\textwidth]{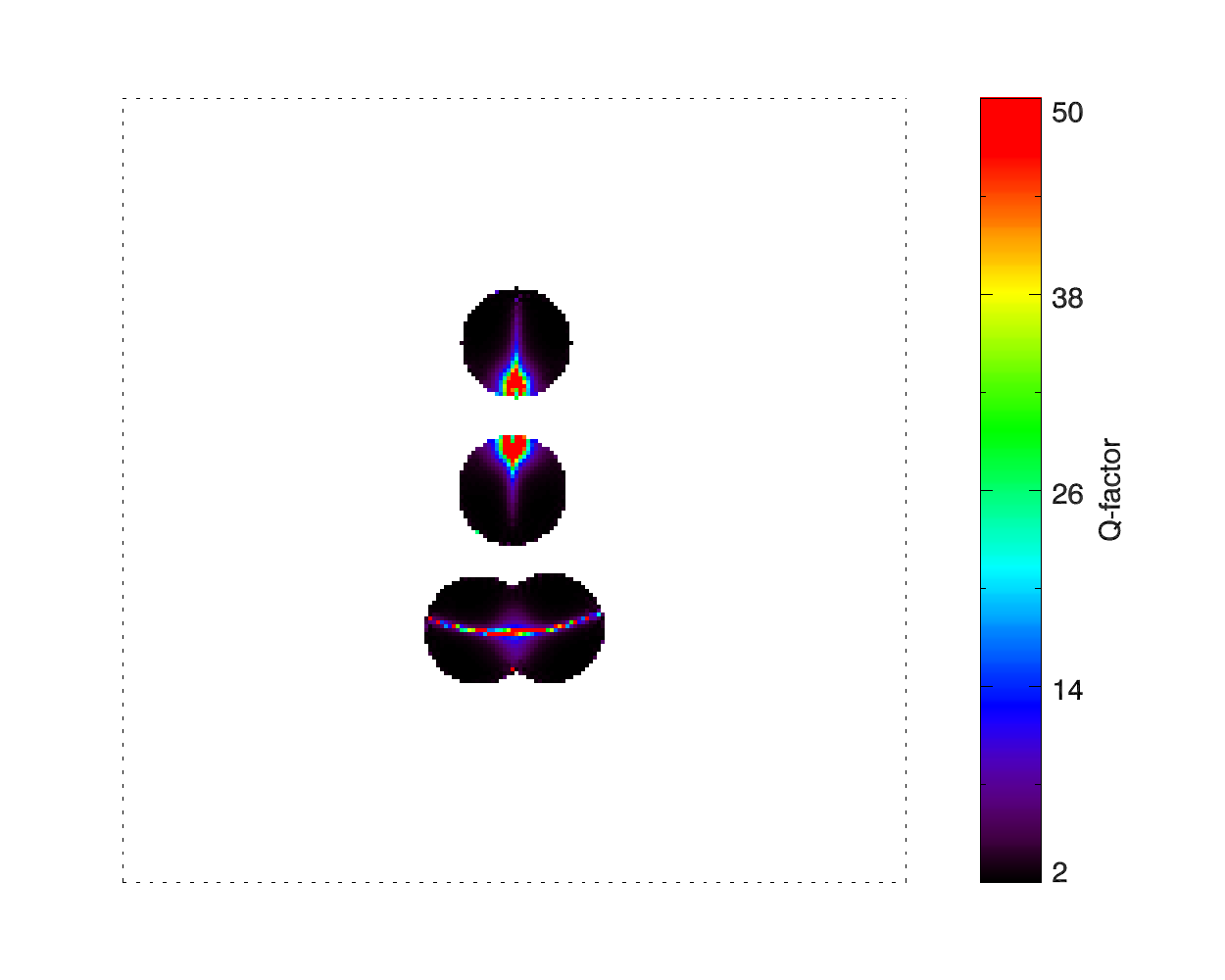}\label{fig:4sposover}} 
    \hfill
    \subfloat[]{\includegraphics[width=0.5\textwidth]{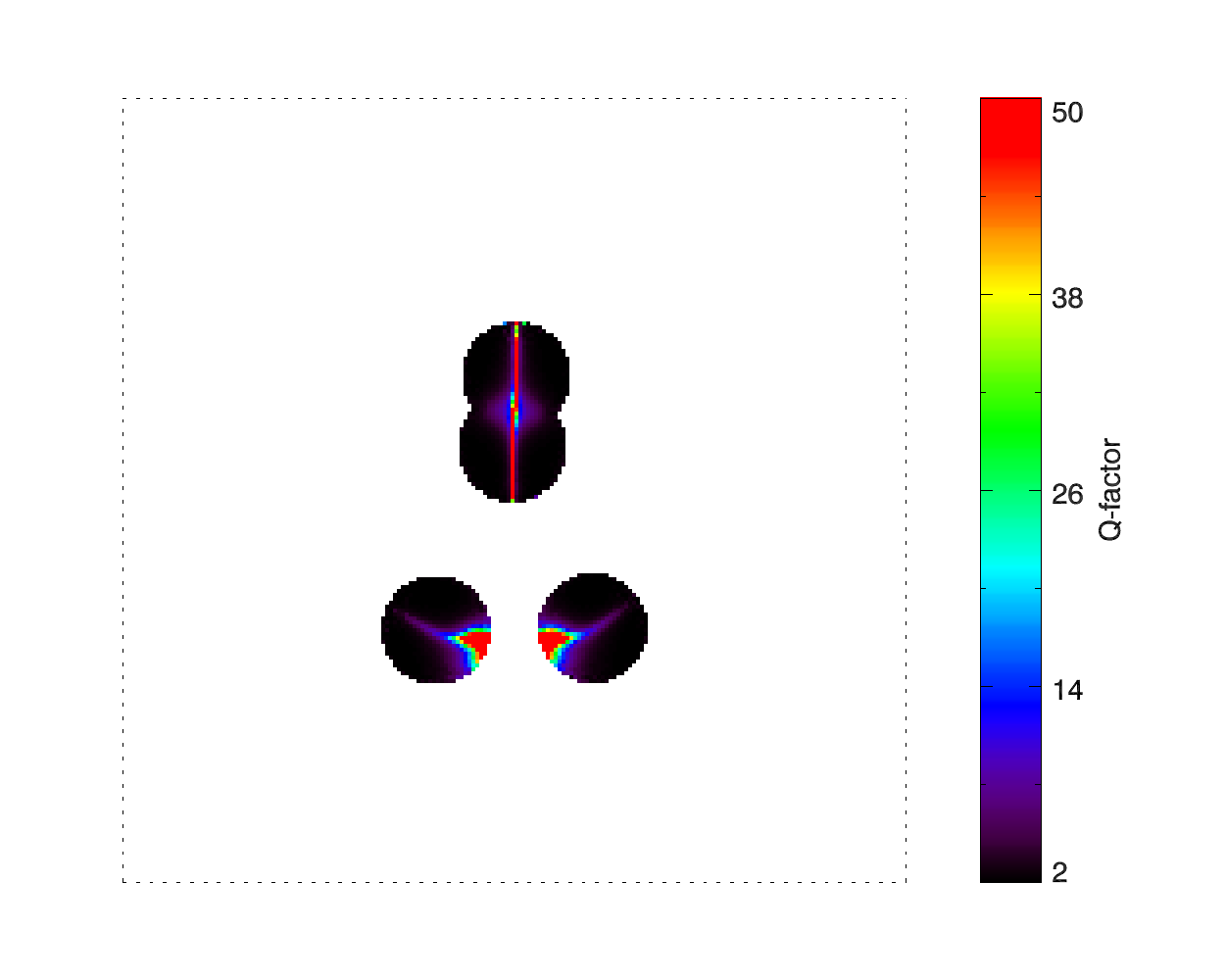}\label{fig:4snegover}}
    \hfill
    \subfloat[]{\includegraphics[width=0.5\textwidth]{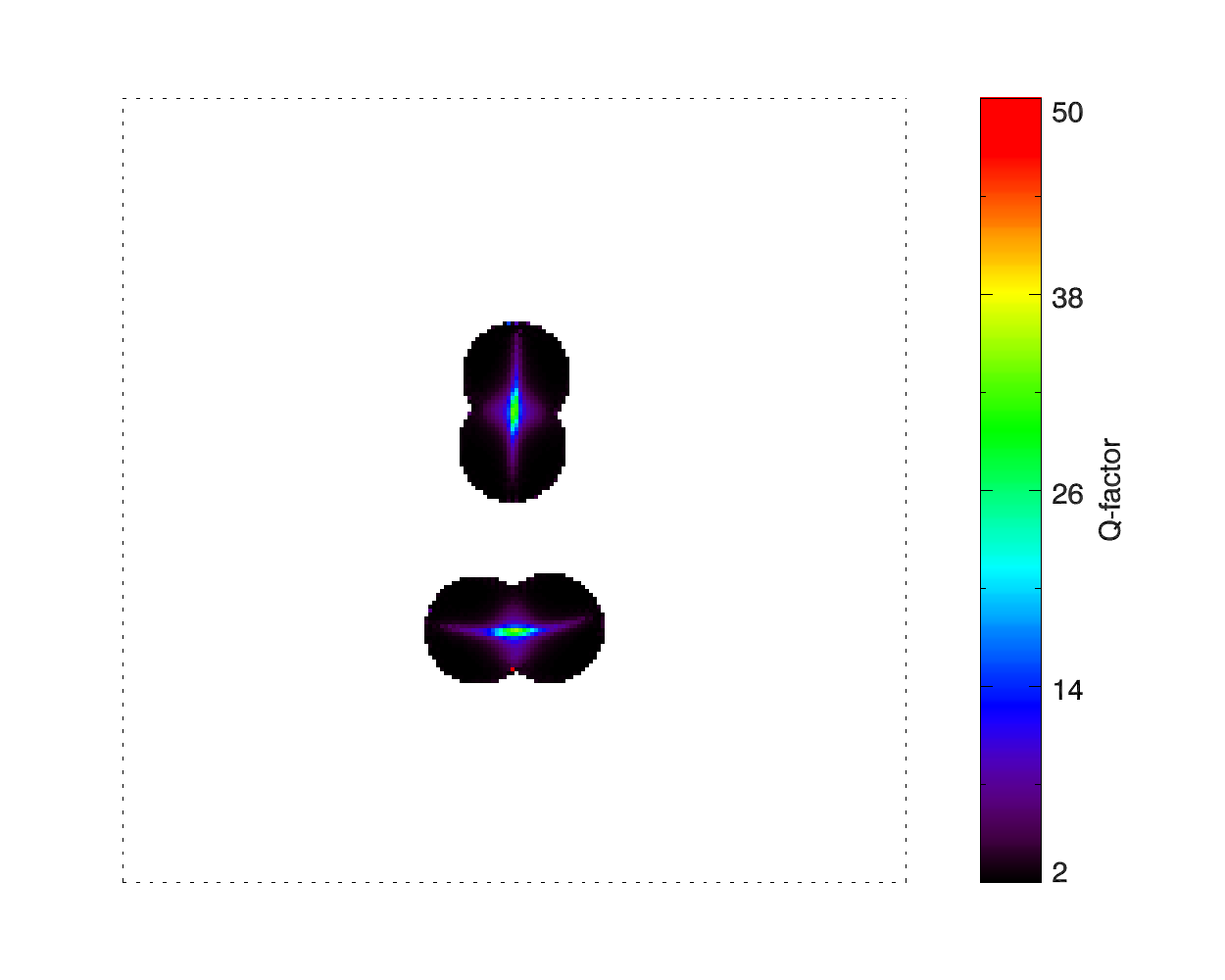}\label{fig:4sbothover}} 
    \caption{$Q$-maps for the four-source case when a) no sources overlap, b) the positive source pair overlap, c) the negative source pair overlap, and d) both source pairs overlap.} \label{fig:fourqs}
    \end{figure}

    The squashing factor is calculated from four field lines per grid cell which has $B_z \neq 0$. It is repeated for all source cell elements to produce a $Q$-map showing how $Q$ varies across the sources. First for the non-overlapping case (Figure \ref{fig:4snoover}), using sources S1 - S4 from Table \ref{tab:4sconfigs}, two distinctive features are present on the `$Q$-maps' for the configuration. The first is a diffuse region where the value of $Q$ gradually increases to a peak, located on each source where it is closest to the other source of the same polarity, whilst the other feature presents as a distinctive line, often only one grid cell across, of a sudden increase in the squashing factor. The diffuse regions indicate the presence of a magnetic null point, with field lines starting close to the null point diverging from one another more than those which originate from cells further away. However, the $Q$-map does not map the null point itself, as this is not located within the boundary of a source. The peaks in $Q$ coincide with the locations where the separatrices extending from the null points meet the photospheric plane. Where the surfaces bisect grid cells, the field lines generated at the cells exist in different domains of connectivity and thus will have a greater divergence than field lines in the same connectivity domain. Thus, the presence of both the null points and their associated surfaces are clearly distinguishable in the non-overlapping case in Figure \ref{fig:4snoover}.

    Next the $Q$-map is generated for the configuration where the (lower) positive sources overlap (Figure \ref{fig:4sposover}), using sources S1$_O$, S2$_O$, S3 and S4 from Table \ref{tab:4sconfigs}. Here, an NLP is present in the overlapping region of the two positive sources. Its presence is indicated by the increase in the $Q$ value, though this signature is less intense than that found about a null point. There is also a diffuse region extended about the signature of the null point between the two negative sources. Here, a long tail of increased $Q$ extends away from the null point, across the negative sources. This tail follows the path of the separatrix-like wall, which extends from the NLP, indicating its presence. The signature of the separatrix dome from the null point is present across the negative sources, as a curve of high $Q$ across the sources. 

    The case where the negative sources overlap, but the positive ones do not, is shown in Figure \ref{fig:4snegover}, using sources S1, S2, S3$_O$ and S4$_O$ from Table \ref{tab:4sconfigs}. Here, the presence of a null point between the two positive sources is clearly illustrated, as is the associated separatrix wall, where it connects down to the negative sources. A gradual diffuse region about the overlap of the negative sources now indicates the presence of the NLP in the overlap. The associated separatrix-like surface is seen across the positive sources as a diffuse, curved region, following the path of the dome. 

    Finally, the case where both the source pairs overlap is studied for completeness (Figure \ref{fig:4sbothover}), using sources S1$_O$ - S4$_O$ from Table \ref{tab:4sconfigs}. Here, the $Q$-map supports the presence of both NLPs and their associated separatrices which are indicated by the diffuse regions across the sources.

    The magnetic skeleton of the final two cases (negative overlap and both polarity overlap) strongly resembles the previous skeletons shown in Figures \ref{fig:4ctopa} and \ref{fig:4SNlP}, with one dome and one wall present in each. When only the two negative sources overlap one another, a separatrix-like surface forms a dome (rather than a wall), originating from the NLP between the negative sources. A separatrix wall is also present, and in this case originates from the null point between the two positive sources. In the case where both source pairs overlap to form two distinct sources, an NLP is present in the overlapping regions. These NLPs are the origins of two separatrix-like surfaces, forming a wall and a dome. Again, a magnetic separator-like field line traces the intersection of these two separatrix-like surfaces. 

    This was also tested with unbalanced pairs of sources, which found that no significant topological changes occur.
    
    The squashing factor has also been employed by \citet{Restante2009} in the search for QSLs, who focused on the transition from the magnetic skeleton to QSLs when magnetic sources are placed below the photospheric plane. They note two categories of locations where they identify QSLs. The first category are those located close to the spine field lines of a sub-photospheric null point, whilst the second category (termed branches) are located at curves defined by the photospheric endpoints of fan field lines which originate from sub-photospheric sources. They find that QSLs associated with spine field lines have maximum $Q$ values close to sub-surface null points, whilst QSLs associated with fan field lines have maximum $Q$ values close to sources. They also pick out another feature of a QSL footprint called a 'null halo', which is attributed as being a diffuse halo region of heightened $Q$ above the location of a sub-surface null point.
    
    Consider again the case presented in this article where both source pairs overlap (Figure \ref{fig:4sbothover}). Here, one can also find heightened values of $Q$ about the NLP position, which appears similar in nature to a null halo. Focusing on the (lower) positive sources, there are then the two elongated regions of increased $Q$ (discussed above) which extend from the possible null halo in the $x$ and $y$ directions. The extension in the $y$ direction has been attributed to the wall separatrix-like surface which originates from the NLP in the (lower) positive source pair. This extension may be analogous to a class of branch identified by \citet{Restante2009}. The extension in the $x$ direction is here attributed to the dome separatrix-like surface originating from the NLP in the (upper) negative source pair. However, following the ideas set out by \citet{Restante2009}, this could actually be two distinct signatures. Suppose that the NLP is located in the region above a sub-surface null point. The central portion of this extended region of heightened $Q$ may be found to match the spine field lines of such a sub-surface null point. Then, the parts of the increased $Q$ extension furthest from the site of the NLP (and so the sub-surface null) may be indicative of a QSL footprint after the spine field line of the sub-surface null has terminated, thus being indicative of a type of branch. The heightened $Q$ signatures presented here may be related to the idea of a sub-surface null and spine, and indeed this article presents similar findings to those of \citet{Restante2009}. However, a direct comparison of the same topological setup would be required to definitively associate separatrix-like surfaces with branches, and to determine if NLPs are located along the x-line structure propagating from a sub-surface null point.

\section{An Open Separatrix Surface}\label{sec:oss}

\begin{figure}[t] 
    \centering
    \subfloat[]{\includegraphics[width=0.5\textwidth,clip=]{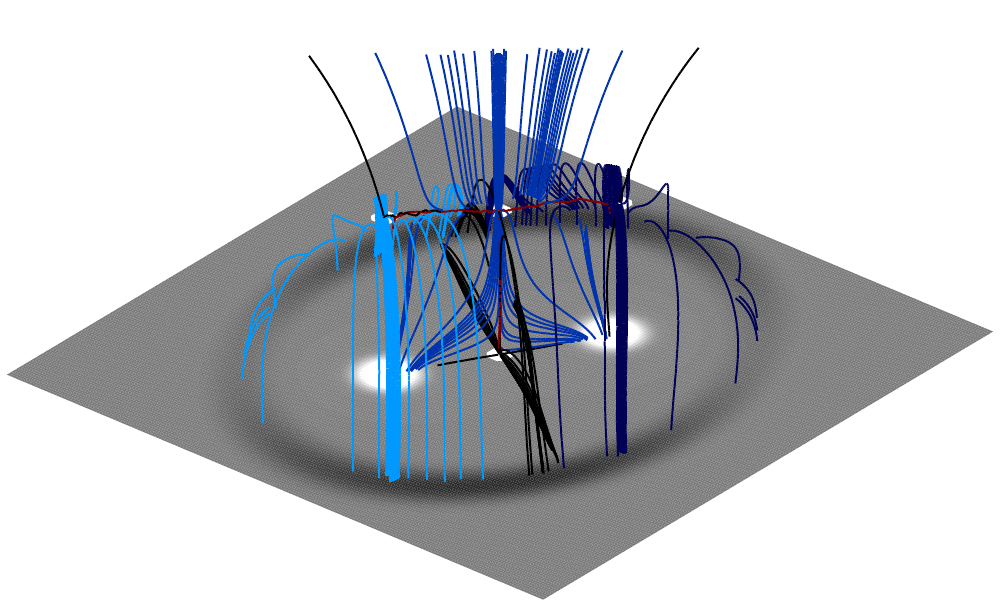}\label{fig:oss}}
    \hfill
    \subfloat[]{\includegraphics[width=0.5\textwidth,clip=]{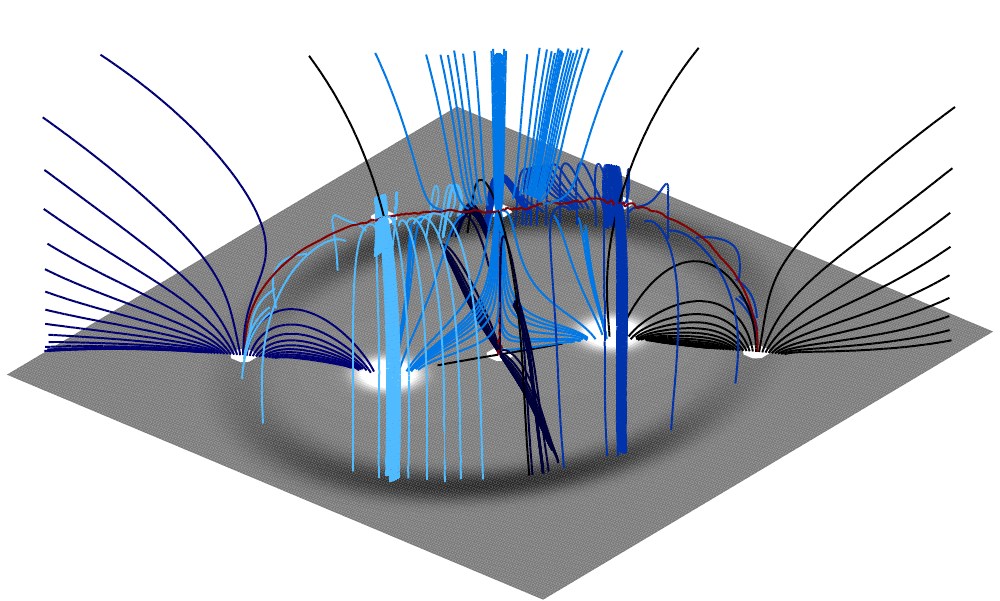}\label{fig:ossnlp}}
    \caption{Topology of the open-separatrix surface case when NLPs are a) excluded from the model and b) included in the model.}\label{fig:ossm}
    \end{figure}

    Next, a series of configurations where two positive sources are contained within the boundary of an annulus of flux is studied. One topological state found has an open separatrix surface, as termed by \citet{Priest2014}, which resembles a heliospheric curtain from the potential field source surface regime \citep{Platten2014} (Figure \ref{fig:oss}). However, the inclusion of NLPs in the analysis of the topology suggests that the surface may not be as open as previously suggested. 

\subsection{The Standard Topological Case}\label{sec:oss2}

    An annulus of negative magnetic flux (-3.0 arbitrary units of flux) encloses two circular sources of positive magnetic flux (0.5 arbitrary units each, producing 1.0 arbitrary units of flux in total). This configuration has an imbalance of flux distributed across the photospheric plane ($z = 0$). For the visualisation of the topology due to this source configuration, boundary conditions follow the description (c) outlined in Section \ref{sec:cspf}, where the normal field for each point on the bounding surfaces is approximated using the analytical expression for the magnetic field. However, when producing the $Q$-maps of the source configurations, the boundary conditions follow description (b), where the flux imbalance is distributed uniformly across the upper plane ($z = n$) bounding the volume. 

    The magnetic skeleton (Figure~\ref{fig:oss}) consists of three coronal null points and a single photospheric null point. Two of the coronal nulls have fan surfaces which project down to the photospheric plane, forming the two coronal domes. The spines of these nulls are upright, extending down to the photosphere, and up into the corona. These spines also bound the fan surface of the `central' coronal null such that it forms the open separatrix surface. The spine field lines of the central coronal null then project down to the photospheric surface, each connecting to the annulus of flux. Together, these spine field lines bound the separatrix wall of the planar null point. While the two appear offset in Figure~\ref{fig:oss}, further investigation shows that the separatrix wall is highly warped near the spines. The fan of the planar null then intersects the fan of the central coronal null in a separator field line, of which there are three in total as the central coronal null points fan also intersects each of the coronal domes. Thus, each null point has one separator field line connecting it to the central coronal null. This topology has three domains of connectivity, two of which are internal to the coronal domes, meaning that the volume exterior to the domes is a single connective domain. This would mean that footpoint motions of a field line may move it from one side of the open separatrix surface to the other, without undergoing reconnection or crossing a separatrix surface. 

    \begin{figure}[ht]
    \centering
    \subfloat[]{\includegraphics[width=0.5\textwidth,clip=]{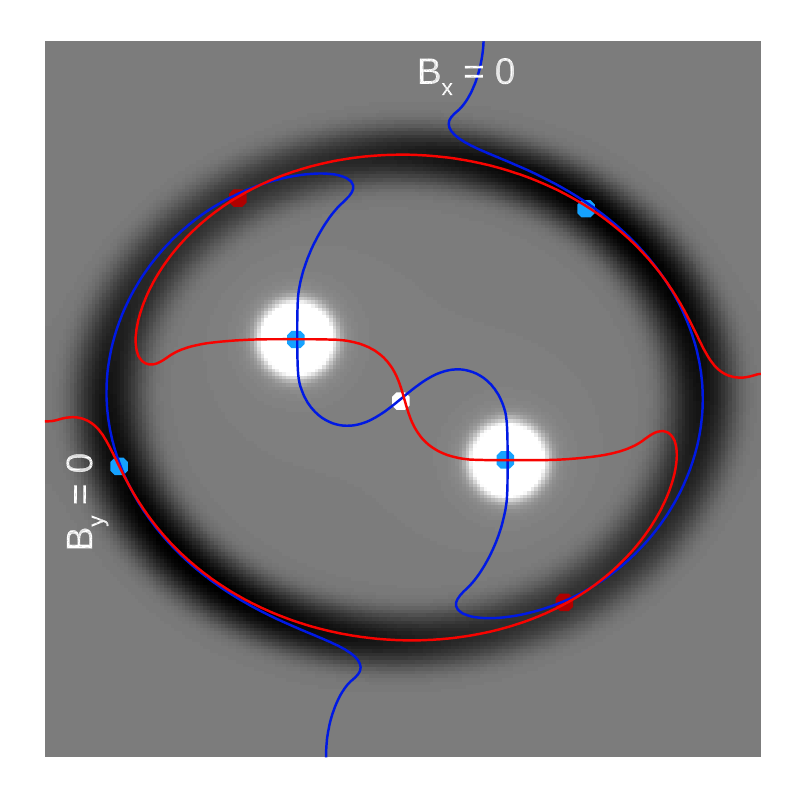}\label{fig:ann_comp}}\\
    \hfill
    \subfloat[]{\includegraphics[width=0.5\textwidth,clip=]{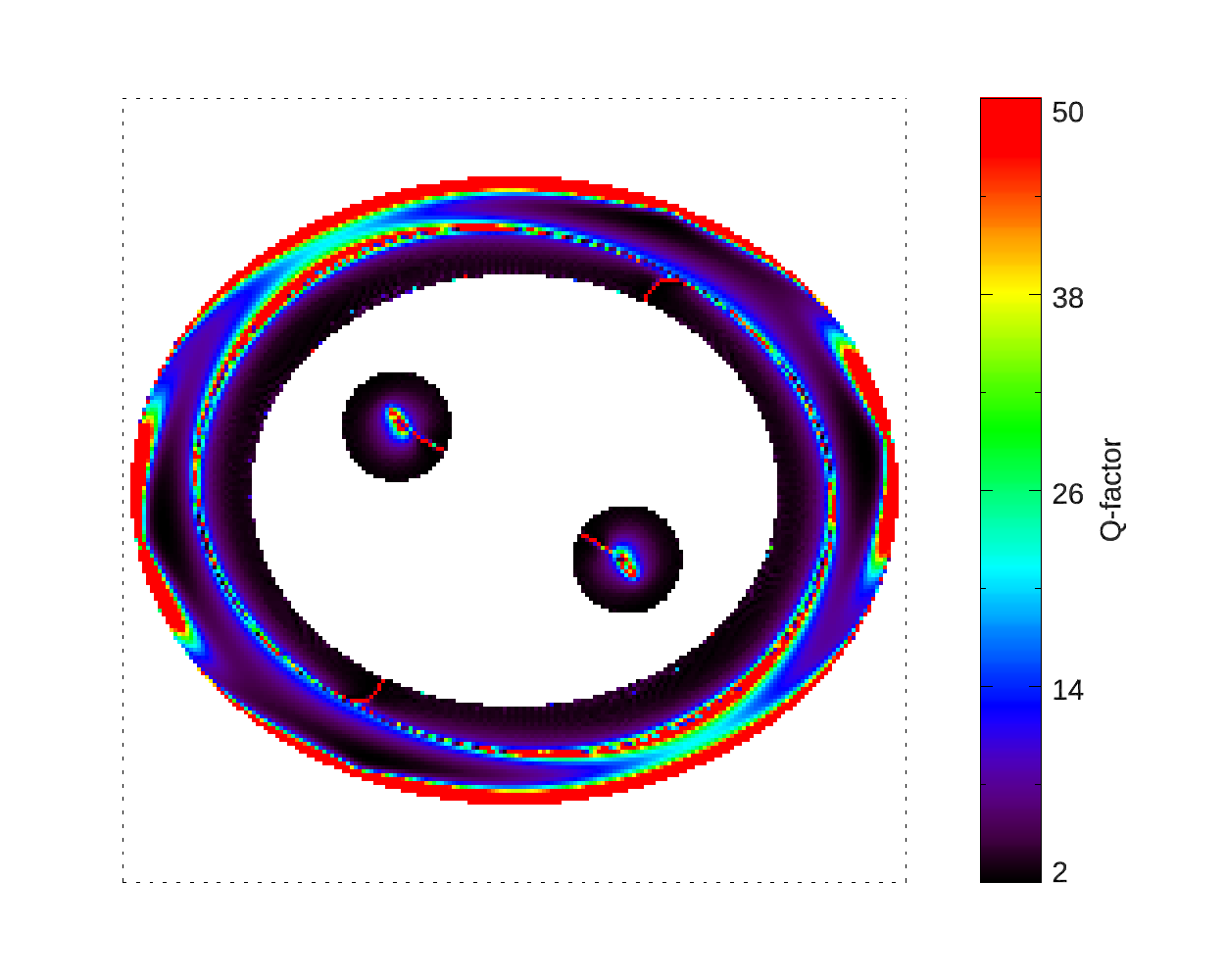}\label{fig:ann_qf}}
    \hfill
    \subfloat[]{\includegraphics[width=0.5\textwidth,clip=]{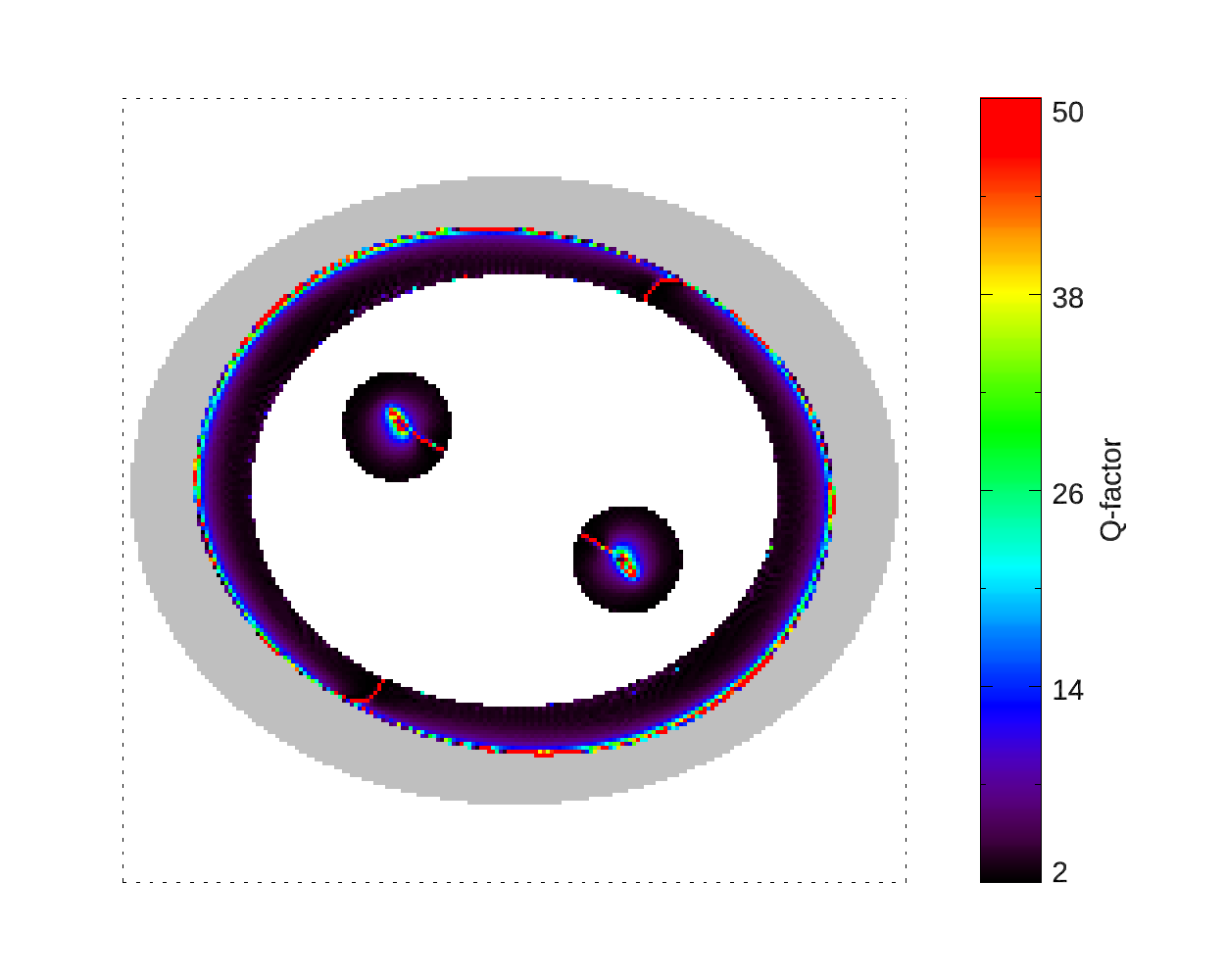}\label{fig:ann_qm}}
    \caption{a) Planar plot of the open-separatrix case. White dots mark null points, red dots are NLPs; and blue dots give SLPs. Nullclines for the configuration are overplotted, blue and red lines are $B_x = 0$ and $B_y = 0$ respectively. b) Complete $Q$-map for the open-separatrix surface case. c) Masked $Q$-map of the source configuration. Where all four test field lines of a grid cell leave the volume, that cell has been shaded grey.}\label{fig:EllPlQm}
    \end{figure}

\subsection{Annulus of flux with Null-like Points}\label{sec:OSSNlps}

    The initial condition and configuration of the sources used in this scenario are the same as those used for the case in Section \ref{sec:oss2}. However, this scenario locates the NLPs and focuses on the changes they introduce to our understanding of the topology. Two NLPs are found within the annulus, along with their fan field lines (Figure \ref{fig:ossnlp}) which define the QSL, in the form of separatrix-like walls. Each of these walls intersects the coronal domes which are present, with a separator-like field line running along each intersection. 
    Figure \ref{fig:ann_comp} shows the positions of the null point, NLPs and SLPs within the source configuration, along with the $B_x$ and $B_y$ nullclines in the region. Mapping the squashing factor for this configuration shows a band of increased $Q$ internal to the source annulus (Figures \ref{fig:EllPlQm} \subref*{fig:ann_qf} and \subref*{fig:ann_qm}). A mask has been applied to the $Q$-map as shown in Figure \ref{fig:ann_qm}. The mask removes cells where the field lines have connected to the $z = n$ surface at the top of the volume, as these are not representative of the field lines connectivities since they do not terminate at another source in the photospheric plane $z = 0$. This band illustrates where the footpoints of the coronal domes meet the $z = 0$ plane which agrees with previously observed results \citep{Masson2009}. At the locations of the NLPs, an increase in $Q$ can be seen in the $Q$-map but this is somewhat masked by the coronal dome marker. The increase in $Q$ due to the separatrix-like surfaces crossing the circular sources is also not as apparent here as it was in Figure \ref{fig:fourqs}. The reasons for diminished signatures in the $Q$-maps are discussed further in Section \ref{sec:sad}.

    When NLPs are considered, the open separatrix surface discussed in Section \ref{sec:oss2} is not as open as previously thought. The motion of a field line from one side of the open separatrix surface to the other would see it cross either the `open separatrix' or a QSL generated by one of the NLPs. Whilst field lines on either side would still be within the same domain of connectivity as one another, motions across the QSL would cause them to undergo slip-running reconnection or magnetic flipping \citep{Aulanier2006, Priest2014}.

\section{Discussion and Conclusion}\label{sec:sad}

    This article has demonstrated that when using constructs from magnetic topology (such as null points, separatrix surfaces, and separators) in a magnetic configuration generated from continuous boundary conditions, care must be taken to ensure that all topological features are included. In particular, x-line structures that originate at null points on the $z=0$ plane in a point-source approximation may not be identified in a continuous regime. X-line structures, such as separators, are important sites in the study of reconnection \citep[e.g.,][]{Longcope1996a, Parnell2008}; if an x-line is present in a magnetic field topology, then it is important to ensure that it is represented appropriately. To address this, the idea of a null-like point (NLP) is introduced. This relaxes the constraint that $B_{z}=0$ at the NLPs to identify topological structures that originate in the $z=0$ plane (or more generally the criteria $\mathbf{B} . \mathbf{n}\ne0$ on an arbitrary boundary is allowed). These give rise to separatrix-like surfaces and the possibility of separator-like structures.
    
    The case studies presented in Section \ref{sec:csset} highlight that the inclusion of NLPs recover topological features lost in the transition from the discrete case and help prevent x-lines being omitted from a topological simulation. These show that the x-line structure coming from the null point between two continuous sources of the same polarity persist when the sources close and overlap. Section \ref{sec:oss} illustrated how even in a truly distributed source, rather than a source overlap (Section \ref{sec:csset}), NLPs can still be identified and provide insight into topological properties that may otherwise be missed. 
    
    In both of these examples, calculation of a squashing factor illustrates that the inclusion of NLPs provides a method for detecting a sub-class of QSL in the magnetic topology. Whilst there will almost certainly be cases where a QSL exists in the absence of an NLP, locating NLPs serves to produce a more complete picture of the topological structure. For example, the open separatrix surface \citep{Priest2014} is bound on each side by the spines of two other coronal null points. For the topologies generated in this article, these spine field lines also form the boundaries for two QSLs originating from the NLPs. Whilst this does not separate the topology into any additional flux domain, it means that additional mechanisms, such as slip-running reconnection, would be required for a field line to cross from one side of the open separatrix surface to the other, without reconnecting across the surface itself. This extends the interpretation of the open separatrix surface given by \citet{Priest2014} and adds key details that would otherwise be missed. 
	
	The work presented here uses potential magnetic fields in both the discrete and continuous situations. However, much like a null point, the definition of a null-like point is not specific to the potential case and may be defined in other regimes such as force-free magnetic field extrapolations and full MHD simulations. This may provide an additional mechanism for the detection of topological features such as quasi-separatrix layers in these approaches.
	

 \begin{acks}
    The authors would like to acknowledge the helpful comments provided by the referee, and the University of Central Lancashire for funding this work.
 \end{acks}
 
 \begin{dpcis}
    The authors declare that there are no conflicts of interest.
 \end{dpcis}

\bibliographystyle{spr-mp-sola}
\bibliography{master_bib}  
%
%
%
\end{article} 
\end{document}